\documentclass{article} 
\usepackage{iclr2020_conference,times}
\usepackage{graphicx,booktabs, float, scrhack}
\usepackage{subcaption}

\usepackage{amsmath,amsfonts,bm}

\def\eqref#1{equation~\ref{#1}}

\def\1{\bm{1}}

\DeclareMathAlphabet{\mathsfit}{\encodingdefault}{\sfdefault}{m}{sl}
\SetMathAlphabet{\mathsfit}{bold}{\encodingdefault}{\sfdefault}{bx}{n}

\usepackage{hyperref}
\usepackage{url}

\title{MODELING, VISUALIZATION, AND ANALYSIS \\ OF AFRICAN INNOVATION PERFORMANCE}

\author{Muhammad Omer, Moayad El-Amin, Ammar Nasr \& Rami Ahmed \thanks{Work was completed while working at Innovation Baylasan, a social enterprise in Khartoum.} \\
Faculty of Engineering \\
University of Khartoum\\
Khartoum, Gamaa ave., Sudan \\
\texttt{\{mhmdgaffar98,mo2yd99,ammarnasraza,ramisketcher\}@gmail.com} \\
}

\iclrfinalcopy 
\begin{document}

\maketitle

\begin{abstract}
In this paper we discuss the concepts and emergence of Innovation Performance, and how to quantify it, primarily working with data from the Global Innovation Index, with emphasis on the African Innovation Performance. We briefly overview existing literature on using machine learning for modeling innovation performance, and use simple machine learning techniques, to analyze and predict the "Mobile App Creation Indicator" from the Global Innovation Index, by using insights from the stack-overflow developers survey. Also, we build and compare models to predict the Innovation Output Sub-index, also from the Global Innovation Index.  
\end{abstract}

\section{Introduction}
Measuring innovation of a certain nation or region emerged as a common field of study in the few last years; as the levels of complexity of each nation and the existing policy structure and infrastructure prove a major challenge in creating a standardized system that can be generalized over regions or over the globe. 
National innovation performance measures do exist in part of the world to pass policies and to explore and come up with improvements in the infrastructure and policy towards innovation in that country. One was the report created by the Advisory Committee of Measuring Innovation in the 21st century economy and presented to the US Secretary of Commerce in 2008, which \_ among-st other goals \_ aimed to develop better ways to quantify innovation in the marketplace and to guide the government towards creating frameworks for measuring innovation and direct the policies that aim to uplift innovation performance in the united states.
Work on regional innovation metrics was a bit more successful. The European Union publishes a yearly report that summarizes the innovative performance of the region called "The Regional Innovation Scoreboard". This scoreboard divided the EU regions into four distinct classes, Regional Innovation Leaders, Regional Strong Innovators, Regional Moderate Innovators, and Regional Modest Innovators. 
Finding a global unified index is an especially difficult task. Where for example pioneering work by \cite{Porter}; on the National Innovative Capacity provides a good historical overview on the countries it covers; the diversity of these countries is the problem as it only takes a look at a specific sector of the world mainly the U.S, Europe and the high to upper-middle income Asian countries with no representation of third world countries and the African region to be exact. 
The Global Innovation Index, which we have chosen as a benchmark for our analysis in this paper, helps to create an environment in which innovation factors are continually evaluated. It is divided into two major indices; the Innovation Input sub-index and the Innovation output sub-index. The innovation input sub-index has five factor evaluating institutions, Human Capital and research, Infrastructure, Market Sophistication and Business sophistication while the Innovation output sub-index gives insights on knowledge and technology outputs and the creative output. The GII solves the problem of representation as it covers 129 countries from different income classes in its 2019 report.
However, the problem for quantifying innovation performance worldwide, and in particular, Africa, is far from solved. The Global Innovation Index is mutilated here and there by missing data and \_ arguably \_ misinforming indicators. Where most of country profiles incomplete come from developing countries specially Africa, we take on one indicator \_ Mobile App Creation \_ that has a high ratio of missing data from the observations in the very same 2019 report, and try to build a robust estimator to interpolate over its values, as we will discuss in the next section.   
Finally, employing machine learning techniques in the analysis and modeling of such innovation performance measures is a promising approach with respect to building robust predictors and providing deep insights that might lead to novel ideas. That being said, however, contributions from the machine learning community with respect to this topic are very few, not to mention that none of them addresses Africa in particular. Two works in particular we think should be addressed are by \citep{David} and \cite{Hajek}, repectively. The first, which we follow a similar approach to it here, performs their analysis on the Global Innovation Index, and the second multi-output artificial neural networks to model regional European innovation, operating on data from the EU's "Nomenclature of Territorial Units for Statistics". Both of these works conclude with the remark that machine learning algorithms perform better on modeling innovation data than traditional analytical methods popular in the literature. 
.
\section{Methodology}
Our two main contributions in this paper are building a model that uses insights from alternative data sources to predict the Mobile App Creation indicator in the Global Innovation Index worldwide data for 2019, and building a model to predict the Innovation Output Sub-Index for a set African countries over the last six years (2014 ~ 2019), also using the Global Innovation Index data for the indicated years. 
For the first one, we used survey data provided by stack-overflow in their website. The survey is organized by stack-overflow, and over 90,000 developers participate in the survey each year, it provides for us deep and comprehensive insights on the state of software development in every country in the world. The survey, which in the form of multiple choice questions, has more than 300 questions and takes about 20 minutes to complete. We have chosen a set of 30 questions from this survey, we believe best represent the status quo of each country's local software market, as well as core competencies for each developer along with their corresponding countries. We then removed unique IDs and averaged one hot encoded column values over the countries, to produce a structure similar to the Global Innovation Index', and finally merged the data set with the Mobile App Creation indicator data, with respect to countries. After that, a correlation matrix was produced to learn about the interaction between features, and determine whether the developer survey's data was relevant at all to the Mobile App Creation Indicator. Finally, we built four models; a Gaussian Process model, an extremely Gradient Boosted Trees (XGBoost) model, Support Vector Machine (SVM) model, and a Random Forests model, and compared their performance using using k-fold cross-validation with k = 10, and with the root mean squared error (RMSE) as a measure. The models accepted all the averaged/weighted survey questions and predicted the Mobile App Creation Indicator.
For the second contribution, we followed an approach similar to the one employed in \cite{David}. We We used the extended report which has 81 detailed features of the 7 aforementioned indices. We extracted the data for Africa for the last six years spanning from 2014 until 2019 and covering 36 African countries with some only being represented in 2 years. Unlike \citep{David}; we didn't drop out countries with incomplete profiles, since that would only end up in eliminating half of the African countries from the data set, also, our analysis focuses on only African countries, with keen consideration to the context and aspects of the shortcomings in developing countries, rather than merely building a predictive model for global innovation performance. We as well, produced a correlation matrix, and employed the four models mentioned above in this analysis as well, the only difference being that this time, the models accepts the 81 features described, and outputs a prediction of the innovation output sub-index. 
Finally, we visualized the fluctuation in innovation performance through these years on a geographical map, using geopandas.

\section{Results}

For the Mobile App Creation prediction, it was found that the data collected from the surveys as weakly correlated with the indicator, which has resulted in the weak performance of the models, except for the Gaussian Processes model which performed best performed best. These poor results, come from the fact that the Global Innovation Index employs a top-bottom approach in the way it collects its data to compute the indicators. For the Mobile App Creation case in particular, it uses "Global downloads of mobile apps, by origin of the HQ firm, scaled by PPP\$ DGP (billions)". This approach explicitly eliminates local innovators, specially those in developing countries who mostly exist as freelancers, social enterprises, or early stage start-ups. This calls for a grassroots movement that extends beyond merely assuming better quantifiers for innovation performance, as well as the pressing need to use alternative data sources to have accurate estimates for quantifying Innovation Performance in the future, specially in the developing world.

\begin{figure}[H]
	\centering
	\begin{subfigure}[t]{5cm}
		\centering
		\includegraphics[width=5cm,height=5cm]{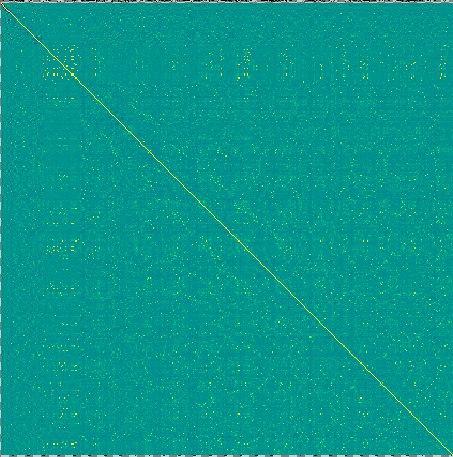}
		\caption{Correlation Matrix for Developer's Survey}
	\end{subfigure}
	\quad
	\begin{subfigure}[t]{5cm}
		\centering
		\includegraphics[width=5cm, height=5cm]{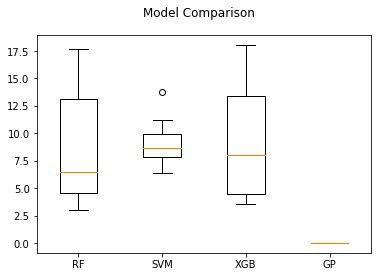}
		\caption{Model Comparison on Developer's Survey Data}
	\end{subfigure}
	\caption{Developer's Survey Data Analysis}
\end{figure}

As for the African countries data set, output of the correlation matrix showed that there is a direct positive correlation between Regulatory Quality of a country and its Innovation output sub-index, meaning an active role of government can lead to a healthier innovative environment and therefore better innovation overall. Another positive correlation appeared between the Innovation Output sub index and Government’s online service and also with Rule of law continuing the apparently needed state sponsoring of innovative friendly policies for acceleration to show in innovation. One of the most interesting correlation was that ISO 14001, an environment certificate appeared in very high correlation with the Innovation output which means that a moving towards a clean environment could push Africa's innovation performance further. Finally the models performed relatively very good, with XGBoost being the dominant with no surprise.

\begin{figure}[H]
	\centering
	\begin{subfigure}[t]{5cm}
		\centering
		\includegraphics[width=5cm,height=5cm]{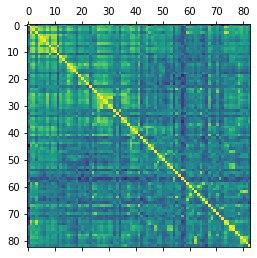}
		\caption{Correlation Matrix for Global Innovation Index Indicators}
	\end{subfigure}
	\quad
	\begin{subfigure}[t]{5cm}
		\centering
		\includegraphics[width=5cm, height=5cm]{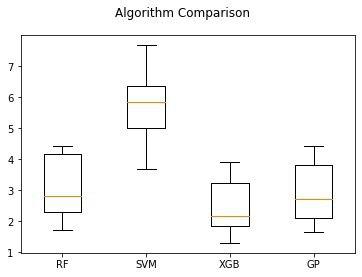}
		\caption{Model Comparison on Global Innovation Index Indicators}
	\end{subfigure}
	\caption{Global Innovation Index Indicators Analysis}
\end{figure}

\section{Conclusion}
In this paper, we outlined the need to use alternative data sources to better measure innovation performance in the continent, as well as the need for grassroots movements to foster and facilitate innovation in Africa. We also showed how simple machine learning techniques can provide novel insights for this matter in question, and finally, a number of observations considering African innovation performance and recommendations to better facilitate innovation and creativity in Africa.

\begin{figure}[H]
	\centering
	\includegraphics[width=100mm]{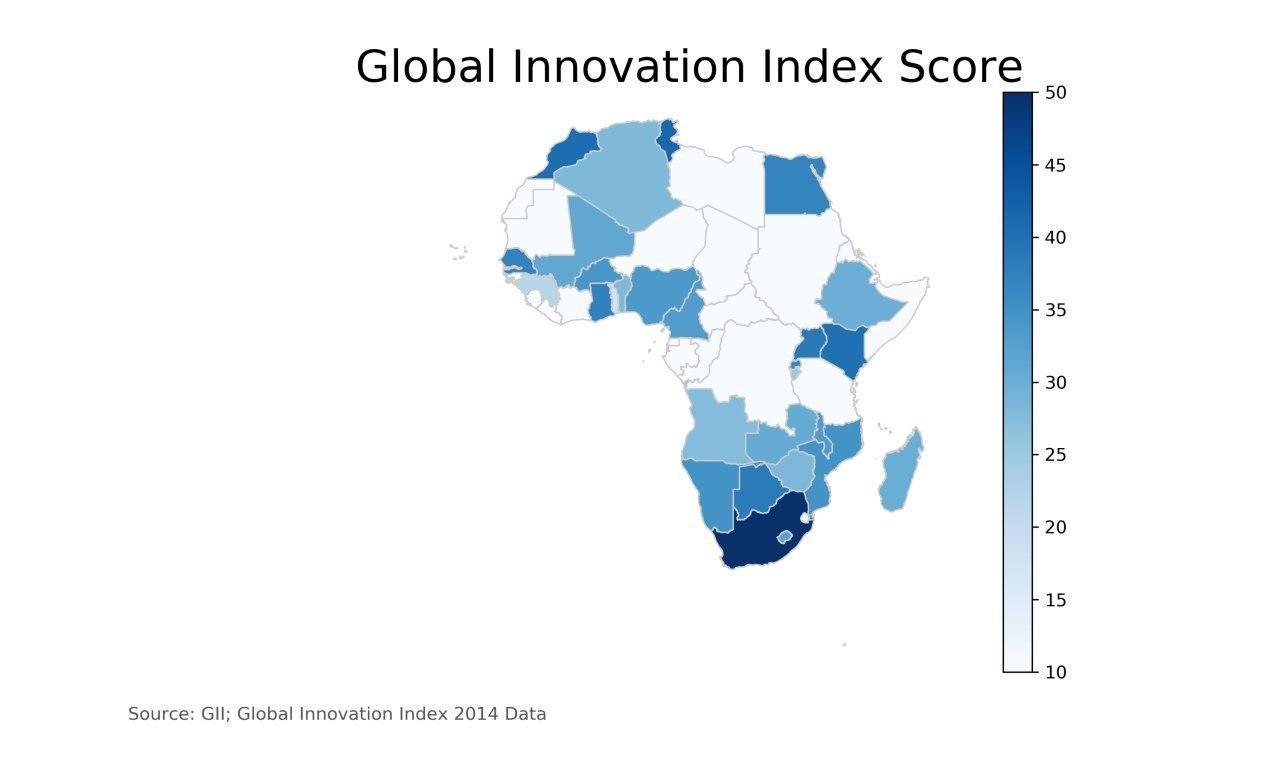}
	\caption{African Innovation Performance in 2014}
	
\end{figure}

\bibliography{iclr2020_conference}

\begin{thebibliography}{5}
\providecommand{\natexlab}[1]{#1}
\providecommand{\url}[1]{\texttt{#1}}
\expandafter\ifx\csname urlstyle\endcsname\relax
  \providecommand{\doi}[1]{doi: #1}\else
  \providecommand{\doi}{doi: \begingroup \urlstyle{rm}\Url}\fi

\bibitem[Bacon et~al.(2019)Bacon, Forner, and Ozcan]{David}
David Bacon, Dominik Forner, and Sercan Ozcan.
\newblock Machine learning approach for national innovation performance data
  analysis.
\newblock In Slimaneand Hammoudi, Christoph Quix, and Jorge Bernardino (eds.),
  \emph{Proceedings of the 8th International Conference on Data Science,
  Technology and Applications}, volume~1. SciTePress, 2019.

\bibitem[Hadzimustafa \& Rexhepi(2011)Hadzimustafa and Rexhepi]{Hadzi}
Shenaj Hadzimustafa and Gadaf Rexhepi.
\newblock Measuring innovation in the 21st century economy.
\newblock \emph{{SSRN} Electronic Journal}, 2011.
\newblock \doi{10.2139/ssrn.1929039}.
\newblock URL \url{https://doi.org/10.2139/ssrn.1929039}.

\bibitem[Hajek \& Henriques(2017)Hajek and Henriques]{Hajek}
Petr Hajek and Roberto Henriques.
\newblock Modelling innovation performance of european regions using
  multi-output neural networks.
\newblock \emph{{PLOS} {ONE}}, 12\penalty0 (10):\penalty0 e0185755, October
  2017.
\newblock \doi{10.1371/journal.pone.0185755}.
\newblock URL \url{https://doi.org/10.1371/journal.pone.0185755}.

\bibitem[Stern et~al.(2000)Stern, Porter, and Furman]{Porter}
Scott Stern, Michael Porter, and Jeffrey Furman.
\newblock The determinants of national innovative capacity.
\newblock Technical report, September 2000.
\newblock URL \url{https://doi.org/10.3386/w7876}.

\bibitem[University et~al.(2019)University, INSEAD, and WIPO]{GII}
Cornell University, INSEAD, and WIPO.
\newblock The global innovation index 2019: Creating healthy lives—the future
  of medical innovation, 2019.
\newblock URL
  \url{https://www.wipo.int/edocs/pubdocs/en/wipo_pub_gii_2019.pdf}.

\end{thebibliography}
\bibliographystyle{iclr2020_conference}

\end{document}